# Twelve years of SAMtools and BCFtools

**Authors:** Petr Danecek, James K. Bonfield, Jennifer Liddle, John Marshall, Valeriu Ohan, Martin O Pollard, Andrew Whitwham, Thomas Keane, Shane A. McCarthy, Robert M. Davies, Heng Li

## Abstract

### Background
SAMtools and BCFtools are widely used programs for processing and analysing high-throughput sequencing data. They include tools for file format conversion and manipulation, sorting, querying, statistics, variant calling and effect analysis amongst other methods.

### Findings
The first version appeared online twelve years ago and has been maintained and further developed ever since, with many new features and improvements added over the years. The SAMtools and BCFtools packages represent a unique collection of tools that have been used in numerous other software projects and countless genomic pipelines.

### Conclusion
Both SAMtools and BCFtools are freely available on GitHub under the permissive MIT licence, free for both non-commercial and commercial use. Both packages have been installed over a million times via Bioconda. The source code and documentation are available from https://www.htslib.org.

## Background

With the advancement of genome sequencing technologies and large scale sequencing projects, new data formats became necessary for interoperability, compact storage and efficient analysis of the data. Among the most common formats used in this field today are SAM [1] and VCF [2] developed by the 1000 Genomes Project [3]. These specialized formats for storing read alignments (SAM) and genetic variants (VCF) are row-oriented tab-delimited text files, which are easy to process using custom scripts but slow to parse and can be inefficient to store. Therefore in practice, the binary counterparts BAM or CRAM are used for alignment data and, when parsing of large VCF files becomes prohibitively slow, BCF provides a more efficient format for processing variation data.

Despite the conceptual simplicity of the underlying DNA sequence, the alignment and variant data carry rich information. This data undergoes a number of processing steps, many of which are algorithmically complex and require specialized software. Also more programming effort and expertise are necessary to parse binary formats. Therefore programs and toolkits that encompass functionality for the most common tasks have been developed. These include tools for file manipulation, quality control, and data analyses, such as sambamba[4], biobambam[5], FastQC[6], and GATK[7]. A successful bioinformatics tool must keep up with advancements in sequencing technologies (e.g. substantial increase in sequencing read length), scale well with ever increasing amounts of data (from single to hundreds of thousands of genomes), expanding focus to encompass new analyses and more complex types of variation. New species also bring challenges such as large chromosomes not representable by 32 bits (over 2 Gbases) or assumptions about the ploidy of an organism. In this article we describe the status, new features and developments in SAMtools and BCFtools.

SAMtools was originally published in 2009[1]. Readers of the online edition of that paper would have been able to download release 0.1.4. The package included not only utilities to convert and manipulate SAM and BAM files, but also a variant caller, which was soon restructured into the BCFtools subpackage (2010, release 0.1.9). Later it became apparent that third-party projects were trying to use code from SAMtools despite it not being designed to be embedded in that way. Therefore the decision was taken in August 2014 (release 1.0) to split the SAMtools package into a standalone library with a well-defined API (HTSlib, currently 82k lines of code), a project for variant calling and manipulation of variant data (BCFtools, 71k lines), and SAMtools for working with alignment data (42k lines). All three projects are maintained in parallel, and improvements to HTSlib naturally filter into new releases of SAMtools and BCFtools. Since the original release the combined size of HTSlib, SAMtools and BCFtools has doubled.

# Findings

## SAMtools

Since the initial release there have been over 2,200 commits to the code repository and 52 releases, the most recent being version 1.11 in September 2020.

The main part of the SAMtools package is a single executable that offers various commands for working on alignment data. The *view* command performs format conversion, file filtering, and extraction of sequence ranges. Files can be reordered, joined and split in various ways using the commands *sort*, *collate*, *merge*, *cat* and *split*. Files can be indexed for fast random access using *index* for alignment files and *faidx* for reference sequences in the FASTA format. File content can be manipulated with commands like *addreplacerg*, *calmd*, *fixmate*, and *reheader*. Duplicated reads, caused by artifacts in the library creation and sequencing process can be flagged using *markdup*. Various statistics on alignment files can be calculated using *idxstats*, *flagstat*, *stats*, *depth* and *bedcov*. Data can be converted to legacy formats using *fasta* and *fastq*. For position-ordered files, the sequence alignment can be viewed using *tview* or output via *mpileup* in a way that can be used for ongoing processing (for example, variant calling). Most recently SAMtools has gained support for amplicon based sequencing projects via *ampliconclip* and *ampliconstats*.

For a complete list of SAMtools commands with a short summary, version and date of the initial commit see Supplement S1. Full documentation for these commands is included with the package in the form of UNIX man pages, and can also be found online at https://www.htslib.org/doc/.

Early releases of SAMtools could read and write alignment data in the SAM and BAM formats. The 1.0 release introduced support for the better-compressed CRAM format [8]. Originally, the program required the use of command-line options to select the input format, and most commands were tied to using BAM files. These restrictions were removed as SAMtools transitioned to use HTSlib, so by release 1.0 most commands could automatically detect the input file format and could directly read and write SAM, BAM and CRAM files. In particular, there is rarely any need to convert SAM to BAM using "*samtools view -b*" before running commands like "*samtools sort*", although regrettably this idiom still appears in a large number of online tutorials. We encourage readers to follow best practices and workflows published at https://www.htslib.org/workflow/.

SAMtools has also become faster, most notably by gaining the ability to use threads to take better advantage of the parallelism available on modern multicore systems. Thread support first arrived in version 0.1.19 (March 2013) which enabled them for sorting and BAM file writing in the *view* command. The number of tasks using threads has slowly increased, so now (thanks to improvements in HTSlib) it is possible to use them for both reading and writing SAM, BAM and CRAM formats in most of the commands. Another time saving improvement is the ability to index files as they are written (added in 1.10). This allows pipelines that need to index files to remove the separate "*samtools index*" stage and associated read-through of the file being indexed.

## BCFtools

The original purpose of the BCFtools package was to divide the I/O and CPU intensive tasks of variant calling into separate steps.

The first step, initially "*samtools mpileup*" but subsequently moved to "*bcftools mpileup*", reads the alignments and for each position of the genome constructs a vertical slice across all reads covering the position ("pileup"). Genotype likelihoods are then calculated, representing how consistent is the observed data with the possible diploid genotypes. The calculation takes into account mapping qualities of the reads, base qualities and the probability of local misalignment, per-base alignment quality (BAQ)[9]. The second step, "*bcftools call*" (known in the initial release as "*bcftools view*"), then evaluates the most likely genotype under the assumption of Hardy-Weinberg equilibrium (in the sample context customizable by the user) using allele frequencies estimated from the data or provided explicitly by the user. In 2016 (release 1.4) genotype likelihood generation was moved from SAMtools to BCFtools to make both variant calling steps part of the same package and to prevent errors arising from the possible use of incompatible versions of the two programs.

Today BCFtools is a full featured program which consists of 21 commands and 38 plugins (single-purpose tools) with more than 230 documented command line switches and options. As of writing, there have been more than 2,300 commits and 29 releases since 2012, with the most recent 1.11 released in September 2020.

The "*bcftools view*" command provides conversion between the text VCF and the binary BCF format, where both formats can be either plain (uncompressed) or block-compressed with BGZF for random access and compact size. The plain text VCF output is useful for visual inspection, for processing with custom scripts, and as a data exchange format. It should not be used when performance is critical, because BCFtools internally uses the binary BCF representation and the conversion between the text VCF format and the binary BCF format is costly. Also compression and decompression is CPU intensive and therefore when streaming between multiple commands in a pipeline it is recommended to stream uncompressed BCF by appending the option "-*Ou*".

The program can do much more than convert between VCF and BCF formats. It can also process third-party formats (using the *convert* command) and manipulate variant files in many ways. It can be used to *index*, *sort* and normalize variants (*norm*), replace headers (*reheader*), add and remove annotations (*annotate*), and subset samples (*view*). Most commands can filter sites either by a region, list of sites, or a general boolean expression involving any combination of VCF tags (--*include*, --*exclude*). Multiple files can be compared, splitting common and file-specific variants into separate files according to custom rules (*isec*). Files sorted by position can also be combined using the *merge* command (input files have different samples) or *concat* command (input files have the same samples). Arbitrary fields can be extracted and formatted into a custom text output (*query*), a feature which, among other things, is useful for scripting.

Apart from file manipulation, the program offers variant callers and algorithms useful for analysis. For calling SNPs and short indels from read alignment files, BCFtools implements two variant calling models. In addition to the original biallelic caller ("*bcftools call -c*"; [10]) there is a newer model available, capable of handling positions with multiple alternate alleles ("*bcftools call -m*") and supporting gVCF output[11]. (For a recent comparison of the variant calling component of the software see [12–15].) The package implements an HMM caller for detection of runs of homozygosity (*roh*; [16]), copy-number variation calling from SNP array data (*cnv*; [17]), and the detection of whole chromosome aberrations (*polysomy*). The program can construct a consensus sequence given a FASTA and a variant file (*consensus*), perform sample identity checks (*gtcheck*) and collect various statistics (*stats*).

In addition to built-in commands, the program supports a dynamic plugin mechanism for specific single-purpose tasks with a diverse range of functions. Examples from a large and ever growing collection include: the plugin *split-vep* for convenient querying and extraction of VEP annotations [18]; *trio-dnm* for ascertainment of *de novo* variants and their parental origin (*parental-origin*), or for collection of statistics (*trio-stats*) in trio data; gVCF manipulation (*gvcfz*); and many more.

For a complete list of BCFtools commands and plugins with a short summary, version and date of the initial commit see Supplement S2. Full documentation for these commands is included with the package in the form of UNIX man pages, and can also be found online together with short tutorials, math notes, and other documentation at https://samtools.github.io/bcftools/.

# Discussion

SAMtools and BCFtools represent a unique collection of tools useful for processing and analysis of sequencing data. Their development has been driven by the need of both large projects and individual user requests issued via GitHub. The code has been installed over a million times via Bioconda[19] and GitHub releases, and more than 900 support and feature requests were resolved on GitHub.

The programs are written in the C programming language and optimized for low memory consumption and high speed. For example, the "*bcftools csq*" command for prediction of functional consequences in a haplotype-aware manner requires only a fraction of the memory required by VEP and is two orders of magnitude faster [20].

Much work has been done to increase the reliability of SAMtools and BCFtools. The test harnesses now include ~700 tests in SAMtools and ~1400 in BCFtools. Continuous integration services run all of the tests on a variety of platforms (including Linux, MacOS and Windows) whenever code is checked into the source repository, ensuring bugs are discovered and fixed rapidly. Code quality is also assured by checking for memory errors, originally using Valgrind memcheck [21] and more recently with AddressSanitizer [22]. Additionally, UndefinedBehaviorSanitizer is used to detect violations of the C standard.

Despite the ever growing sample sizes and rapid increases in the amount of sequenced data, the programs have withstood the test of the time. However, extremely big files are produced by large projects and their processing requires a high degree of parallelization on computing clusters. Future versions of SAMtools and BCFtools are expected to make more use of threaded code to allow faster processing of such files. Sometimes even the limits of BCF representation itself can be reached. For example, highly polymorphic sites can contain dozens of alternate indel alleles which, in files with tens of thousands of samples, exceed the internal limit of 4GB per site due to quadratic scaling of annotations such as FORMAT/PL. An extension of the VCF specification has been proposed to address this problem by introducing a localized version of such annotations with linear scaling[23] and has been implemented in BCFtools.

The programs have been used to process and analyze sequencing data from all types of species: vertebrate, non-vertebrates, pathogens, plants and viruses. This provides interesting challenges and opportunities for future development. For example, some of the BCFtools commands are limited to handling haploid and diploid organisms and the support for large "64-bit" genomes is currently only partial. More work is also planned to overcome difficulties stemming from ambiguities in VCF allele encoding (such as operations of atomization and deatomization), to improve visualization of results, and there are at least 50 feature requests currently registered on GitHub to investigate.

# Availability of supporting source code and requirements

Project name: SAMtools
Project home page: [https://www.htslib.org](https://www.htslib.org),  [https://github.com/samtools/samtools](https://github.com/samtools/samtools)
Operating system(s): Platform independent
Programming language: C
License: MIT/Expat
RRID: SCR_002105
biotools:samtools

Project name: BCFtools
Project home page: [https://www.htslib.org](https://www.htslib.org),  [https://github.com/samtools/bcftools](https://github.com/samtools/bcftools)
Operating system(s): Platform independent
Programming language: C
Other requirements: Optional use of GNU Scientific Library (GSL)
License: MIT/Expat
RRID: SCR_002105
biotools:bcftools

# Additional Files

Additional File, Table S1: Table of SAMtools commands
Additional File, Table S2: Table of BCFtools commands

# Competing Interests

The authors declare they have no competing interests.

# Authors' contributions

J.B., P.D., R.D., H.L., J.M., M.P., V.O., and A.W. wrote the SAMtools software with J.L. and S.M. supporting;  R.D., T.K., and J.M. provided supervision.
P.D. and S.M. wrote the BCFtools software with J.B., R.D., H.L., J.M. and V.O. supporting; S.M. also provided supervision.
J.B., P.D., R.D., V.O., and A.W. wrote the original draft of the manuscript with all authors reviewing.

# Funding

This work was supported by the Wellcome Trust grant [206194].

https://github.com/samtools/hts-specs/pull/434

# Additional material for
# Twelve years of SAMtools and BCFtools

## Table S1. Table of SAMtools commands

List of SAMtools commands with the date of the initial commit, and the version number of the release where the command became available.  The coverage command was contributed by Florian Breitwieser, all others are the work of the authors.  Some commands have been renamed - where this has happened, the first release that accepted the new name is noted.  While the old names are no longer documented, in all cases they will be accepted as an alias for the new name if used.

| Date | Version | Command | Description |
|---|---|---|---|
| 2008-12-22 | 0.1.1 | faidx | index/extract FASTA |
| 2008-12-22 | 0.1.1 | merge | merge sorted alignments |
| 2008-12-22 | 0.1.1 | sort | sort alignment file |
| 2008-12-22 | 0.1.1 | tview | text alignment viewer |
| 2008-12-22 | 0.1.1 | view | SAM<->BAM<->CRAM conversion |
| 2008-12-22 | 0.1.1 | index | index alignment |
| 2009-01-22 | 0.1.2 | fixmate | fix mate information |
| 2009-01-29 | 0.1.3 | flagstat | simple stats |
| 2009-04-24 | 0.1.4 (as fillmd)<br>0.1.6 (as calmd) | calmd | recalculate MD/NM tags and '=' bases |
| 2010-06-11 | 0.1.8 | reheader | replace BAM header |
| 2010-06-12 | 0.1.8 | mpileup | multi-way pileup |
| 2010-06-13 | 0.1.8 | idxstats | BAM index stats |
| 2011-02-25 | 0.1.13 | phase | phase heterozygotes |
| 2011-02-25 | 0.1.13 | targetcut | cut fosmid regions (for fosmid pool only) |
| 2011-03-18 | 0.1.14 | cat | concatenate BAMs |
| 2011-04-01 | 0.1.15 | depth | compute the depth |
| 2012-02-08 | 0.1.19 | depad | convert padded BAM to unpadded BAM |
| 2012-04-17 | 0.1.19 | bedcov | read depth per BED region |
| 2012-05-17 | 0.1.19 (as bamshuf)<br>1.3 (as collate) | collate | shuffle and group alignments by name |
| 2013-09-09 | 1 | stats | generate stats (former bamcheck) |
| 2013-11-28 | 1 | flags | explain BAM flags |
| 2014-03-10 | 1 | split | splits a file by read group |
| 2014-04-17 | 1.0 (as bam2fq)<br>1.3 (as fastq) | fastq | converts a BAM to a FASTQ |
| 2015-03-19 | 1.3 | dict | create a sequence dictionary file |
| 2015-03-30 | 1.3 | addreplacerg | adds or replaces RG tags |
| 2015-06-11 | 1.3 | quickcheck | quickly check if SAM/BAM/CRAM file appears intact |
| 2015-08-24 | 1.3 | fasta | converts a BAM to a FASTA |
| 2017-08-08 | 1.6 | markdup | mark duplicates |
| 2018-05-21 | 1.9 | fqidx | index/extract FASTQ |
| 2018-12-11 | 1.1 | coverage | alignment depth and percent coverage |
| 2020-04-09 | 1.11 | ampliconclip | clip oligos from the end of reads |
| 2020-04-30 | 1.11 | ampliconstats | generate amplicon specific stats |

## Table S2. Table of BCFtools commands

List of BCFtools commands and plugins (prefixed with '+') with the date and version of the initial commit. Marked are contributions by David Laehnemann (*), Nicola Asuni (+) and Giulio Genovese (#).

| Date | Version | Command | Description |
|---|---|---|---|
| 2012-05-17 | 0.1.0 | view | VCF/BCF conversion, view, subset and filtering |
| 2012-08-07 | 0.1.0 | merge | merge VCF/BCF files files from non-overlapping sample sets |
| 2012-09-06 | 0.1.0 | isec | intersections of VCF/BCF files |
| 2013-02-07 | 0.1.0 | query | transform VCF/BCF into user-defined formats |
| 2013-02-12 | 0.1.0 | filter | filter VCF/BCF files using fixed thresholds |
| 2013-03-04 | 0.1.0 | gtcheck | check sample concordance, detect sample swaps and contamination |
| 2013-03-12 | 0.1.0 | norm | left-align and normalize indels, and more |
| 2013-08-20 | 0.1.0 | call | -m and -c calling |
| 2013-08-30 | 0.1.0 | stats | produce VCF/BCF stats |
| 2013-11-05 | 0.2.0 | annotate | annotate and edit VCF/BCF files |
| 2013-11-05 | 0.2.0 | roh | identify runs of autozygosity (HMM) |
| 2014-01-10 | 0.2.0 | +missing2ref | sets missing genotypes ("./.") to ref allele ("0/0" or "0\|0") |
| 2014-01-14 | 0.2.0 | concat | concatenate VCF/BCF files from the same set of samples |
| 2014-02-07 | 0.2.0 | index | index VCF/BCF files |
| 2014-04-11 | 0.2.0 | +counts | minimal plugin which counts number of SNPs, Indels, and total number of sites |
| 2014-04-11 | 0.2.0 | +dosage | prints genotype dosage |
| 2014-04-23 | 0.2.0 | +frameshifts | annotate frameshift indels |
| 2014-07-01 | 0.2.0 | reheader | modify VCF/BCF header, change sample or chromosome names |
| 2014-07-29 | 0.2.0 | convert | convert VCF/BCF files to different formats and back |
| 2014-09-02 | 1.0 | cnv | CNV calling from array data (HMM) |
| 2014-09-03 | 1.0 | polysomy | detect number of chromosomal copies from Illumina's B-allele frequency |
| 2014-09-25 | 1.1 | +fixploidy | sets correct ploidy |
| 2014-10-01 | 1.1 | consensus | create consensus sequence by applying VCF variants |
| 2015-01-08 | 1.1 | +tag2tag | convert between similar tags, such as GL and GP |
| 2015-07-28 | 1.2 | +fill-tags | set various INFO tags |
| 2015-09-16 | 1.2 | +setGT | set genotypes according to rules requested by the user |
| 2015-10-02 | 1.2 | +color-chrs | color shared chromosomal segments, requires trio VCF with phased GTs |
| 2015-10-02 | 1.2 | +impute-info | add imputation information metrics to the INFO field based on selected FORMAT tags |
| 2015-10-02 | 1.2 | +mendelian | count Mendelian consistent / inconsistent genotypes |
| 2016-02-23 | 1.3 | +GTisec (*) | count genotype intersections across all possible sample subsets in a vcf file |
| 2016-06-21 | 1.3.1 | +guess-ploidy | determine sample sex by checking genotype likelihoods (GL,PL) or genotypes (GT) |
| 2016-06-28 | 1.3.1 | +ad-bias | find positions with wildly varying ALT allele frequency (Fisher test on FMT/AD) |
| 2016-06-28 | 1.3.1 | +trio-switch-rate | calculate phase switch rate in trio samples, children samples must have phased GTs |
| 2016-08-02 | 1.3.1 | +af-dist | collect AF deviation stats and GT probability distribution given AF and assuming HWE |
| 2016-08-03 | 1.3.1 | +fixref | determine and fix strand orientation |
| 2016-08-03 | 1.3.1 | +GTsubset (*) | output only sites where the requested samples all exclusively share a genotype |
| 2016-08-05 | 1.3.1 | mpileup | moved from samtools |
| 2016-09-02 | 1.3.1 | csq | call variation consequences |
| 2016-09-21 | 1.3.1 | +fill-from-fasta | fill INFO or REF field based on values in a fasta file |

| Date | Version | Command | Description |
|---|---|---|---|
| 2016-11-23 | 1.3.1 | +isecGT | compare two files and set non-identical genotypes to missing |
| 2017-04-21 | 1.4 | +check-sparsity | print samples without genotypes in a region or chromosome |
| 2017-05-17 | 1.4.1 | +prune | prune sites by missingness, allele frequency or linkage disequilibrium |
| 2017-06-29 | 1.5 | +check-ploidy | check if ploidy of samples is consistent for all sites |
| 2017-07-06 | 1.5 | sort | sort VCF/BCF file |
| 2017-11-12 | 1.6 | +split | split VCF by sample, creating single- or multi-sample VCFs |
| 2018-03-15 | 1.7 | +contrast | simple association test, checks for novel alleles and genotypes in two groups of samples |
| 2018-05-18 | 1.8 | +trio-stats | calculate transmission rate in trio children |
| 2018-06-28 | 1.8 | +smpl-stats | calculates basic per-sample stats |
| 2018-07-04 | 1.8 | +add-variantkey (+) | add VariantKey INFO fields VKX and RSX |
| 2018-07-04 | 1.8 | +allele-length (+) | count the frequency of the length of REF, ALT and REF+ALT |
| 2018-07-04 | 1.8 | +variantkey-hex (+) | generate unsorted VariantKey-RSid index files in hexadecimal format |
| 2018-10-04 | 1.9 | +gvcfz | compress gVCF file by resizing non-variant blocks according to specified criteria |
| 2019-03-06 | 1.9 | +remove-overlaps | remove overlapping variants and duplicate sites |
| 2019-03-25 | 1.9 | +split-vep | extract fields from structured annotations such as INFO/CSQ created by bcftools/csq or VEP |
| 2019-04-24 | 1.9 | +parental-origin | determine parental origin of a CNV region |
| 2019-04-29 | 1.9 | +indel-stats | calculates per-sample or de novo indels stats |
| 2020-06-23 | 1.10.2 | +scatter (#) | intended as an inverse to `bcftools concat`, scatter VCF by chunks or regions, creating multiple VCFs |
| 2020-12-16 | 1.11 | +trio-dnm2 | screen variants for possible de-novo mutations in trios |